%% file: IRS_WCL_final_arxiv.tex
\newif\ifconfver
\confverfalse      

\newif\ifplainver  
  \plainvertrue   

\ifplainver
    \confverfalse   
\fi

\ifconfver
     \documentclass[10pt,twocolumn,twoside]{IEEEtran}
\else
    \ifplainver
        \documentclass[12pt,draftcls,onecolumn]{IEEEtran}
    \else
\documentclass[11pt]{article}
        \usepackage{fullpage}
    \fi
\fi

\usepackage{calc,amsfonts,amssymb,amsmath,bm,url,color,theorem,graphicx,cite}
\usepackage{enumerate}
\usepackage{psfrag,float}
\usepackage{algorithm}
\usepackage{array}
\usepackage{algorithmic}
\usepackage{subcaption,epstopdf}
\usepackage{xr}
\usepackage{soul}
\usepackage{multirow}

\definecolor{orange}{RGB}{255,107,0}

\theorembodyfont{\rmfamily}

\newcommand\bx{\ensuremath{{\bm x}}}
\newcommand\by{\ensuremath{{\bm y}}}
\newcommand\bG{\ensuremath{{\bm G}}}
\newcommand\bh{\ensuremath{{\bm h}}}
\newcommand\bH{\ensuremath{{\bm H}}}

\newcommand\bz{\ensuremath{{\bm z}}}

\newcommand\bX{\ensuremath{{\bm X}}}
\newcommand\bZ{\ensuremath{{\bm Z}}}

\newcommand\bd{\ensuremath{{\bm d}}}

\newcommand{\Cbb}{\mathbb{C}}

\newcommand{\setX}{\mathcal{X}}

\newcommand{\setS}{\mathcal{S}}

\newcommand{\setD}{\mathcal{D}}

\newcommand{\jj}{\mathfrak{j}}
\newcommand{\dec}{\mathrm{dec}}

\newcolumntype{M}[1]{>{\centering\arraybackslash}m{#1}}

\makeatletter
\def\bstctlcite{\@ifnextchar[{\@bstctlcite}{\@bstctlcite[@auxout]}}
\def\@bstctlcite[#1]#2{\@bsphack
  \@for\@citeb:=#2\do{%
    \edef\@citeb{\expandafter\@firstofone\@citeb}%
    \if@filesw\immediate\write\csname #1\endcsname{\string\citation{\@citeb}}\fi}%
  \@esphack}
\makeatother
\begin{document}

\bibliographystyle{IEEEtran}

\newcommand{\papertitle}{
 Minimum Symbol-Error Probability
 Symbol-Level Precoding with
Intelligent Reflecting Surface}

\newcommand{\paperabstract}{
Recently, the use of intelligent reflecting  surface (IRS) has gained considerable attention in wireless communications.
By intelligently adjusting  the passive reflection angle,  IRS is able to assist the base station (BS) to  extend the coverage and improve spectral efficiency. This paper considers a joint symbol-level precoding (SLP) and IRS reflecting design to minimize the symbol-error probability (SEP) of the intended users in an IRS-aided multiuser MISO downlink. We formulate the SEP minimization problems to pursue uniformly good performance for all users for both QAM and PSK constellations.
The resulting problem is non-convex and we resort to alternating minimization to obtain a stationary solution.
Simulation results demonstrate that under the aid of IRS our proposed design indeed enhances the bit-error rate performance. In particular, the performance improvement is significant when the number of IRS elements is large.
}


\ifplainver


    \title{\papertitle}

    \author{
Mingjie Shao, Qiang Li, and Wing-Kin Ma
\thanks{ M. Shao and W.-K. Ma are with the Department of Electronic Engineering, The Chinese University of Hong Kong, Hong Kong (e-mail:
mjshao@ee.cuhk.edu.hk; wkma@cuhk.edu.hk).}
\thanks{Q. Li ({\it corresponding author}) is with the School of Information and Communication Engineering,
University of Electronic Science and Technology of China, Chengdu 611731,
China (e-mail: lq@uestc.edu.cn).    }
}

    \maketitle
    \begin{abstract}
    \paperabstract
    \end{abstract}

        \begin{IEEEkeywords}\vspace{-0.0cm}
        intelligent reflecting surface, symbol-level precoding, symbol-error probability
    \end{IEEEkeywords}


\else
    \title{\papertitle}

    \ifconfver \else {\linespread{1.1} \rm \fi

    \author{Mingjie Shao, Qiang Li, and Wing-Kin Ma}

    \maketitle

    \ifconfver \else
        \begin{center} \vspace*{-2\baselineskip}
        \end{center}
    \fi

    \begin{abstract}
    \paperabstract
    \end{abstract}

    \begin{IEEEkeywords}\vspace{-0.0cm}
        intelligent reflecting surface, symbol-level precoding, symbol-error probability
    \end{IEEEkeywords}

    \ifconfver \else \IEEEpeerreviewmaketitle} \fi

 \fi

%
\section{Introduction}

Recently, the intelligent reflecting  surface (IRS) is proposed to provide new degrees of freedom to enhance the performance of wireless communication systems \cite{huang2019holographic}.
IRS is a programmable passive reflecting array. By collaborating with the base station (BS), the reflecting angle of the IRS elements are adjusted to create better propagation conditions for the intended users, which leads to benefits such as enhanced transmission quality, extended  coverage, etc.
However, the emergence of IRS also raises a problem: How can we integrate  IRS in current wireless communication systems such that the effect of the reflecting signals are constructive for the intended users?

The fast growing body of literature has considered designing the IRS under various system performance metrics.
The single-user case has been well studied in \cite{wu2019intelligent,abeywickrama2019intelligent}.
The multiuser case has also been investigated by joint precoding and IRS design.
By properly choosing the reflecting angles of the IRS elements, the existing works have shown that the IRS-aided systems are able to reduce transmission power at the BS \cite{huang2019reconfigurable,wu2019intelligent}, improve the weighted sum-rate \cite{guo2019weighted,jung2020performance}, enhance information security \cite{cui2019secure}, to name a few.
More recently, practical IRS implementations  for  discrete reflecting phase restriction \cite{wu2019beamforming} and joint design with reflecting amplitude \cite{abeywickrama2020intelligent} 
 have gained increasing attention. 
Up to now, most of the existing works for IRS-aided communication systems are built upon conventional linear precoding schemes, while the more advanced symbol-level precoding (SLP) schemes have not been well studied under the IRS.
The salient feature of SLP in comparison with conventional linear precoding is that SLP takes advantage of specific symbol constellation structure (e.g., PSK and QAM) in the design for enhanced symbol-error probability (SEP) performance~\cite{masouros2015exploiting,liu2018symbol}.
A concurrent IRS work considering the SLP design for PSK and for power minimization is proposed very recently \cite{liu2019joint}.

This paper considers a joint SLP and IRS design for minimizing the SEP of all the users in an IRS-aided multiuser MISO downlink and for both QAM and PSK constellations.
Our design is built on a widely adopted IRS model, where the reflecting coefficient is assumed to have the largest reflecting amplitude and continuous phase \cite{huang2019reconfigurable,nayeri2018reflectarray,wu2019intelligent,guo2019weighted,cui2019secure}. 
Under the total power constraint at the BS, a worst-case SEP minimization problem is formulated.  Our minimum SEP-based design formulation follows that of our very recent work for one-bit and constant-envelope MIMO precoding \cite{shao2019framework}, where the harsh signal constraints were handled by a custom-built penalty method; in this paper we analyze the SEP under the aid of IRS and exploit the constellation structure to produce a tighter SEP upper bound than that in \cite{shao2019framework}. 
The resulting problem is non-smooth and non-convex.
We first apply smooth approximation to the objective function to obtain a smooth problem.
Then, we alternately optimize the precoder and IRS reflecting angles by the  accelerated projected gradient (APG) method for each subproblem. 
 Numerically, the proposed algorithm exhibits fast convergence rate.
Simulation results show that the IRS-aided SLP design is able to boost the BER performance compared to that without the aid of IRS.

\section{Problem Formulation}


We consider an IRS-aided multiuser MISO downlink transmission.
A BS with $N$ transmit antennas transmits  data streams to $K$ single-antenna users simultaneously with the aid of an IRS.
The IRS is equipped with $M$ independent passive reflecting elements and with reflecting coefficients $\bm\theta\in \Cbb^{M}$,
which can be modeled by
\[
  \theta_i =\beta_i\cdot e^{\jj\phi_i},~i=1,\ldots, M,
\]
where $\beta_i\in [0,1]$ is the complex gain and $\phi_i\in [0,2\pi)$ is the reflecting angle. 
In practice, the reflecting coefficient $\bm\theta$ is commonly designed to have the largest amplitude and to be continuous, i.e., $|\theta_i|=1$, for simplicity \cite{huang2019reconfigurable,nayeri2018reflectarray,wu2019intelligent,guo2019weighted,cui2019secure}, which is also adopted throughout this paper.
The channel from the BS to the IRS, the channel from the BS to user $i$, and the channel from the IRS to user $i$ are denoted by $\bG$, $\bh_{d,i}$ and $\bh_{r,i}$, respectively.
Assuming frequency-flat  block fading channels, the received signal at user side is given by
\begin{equation}\label{eq:IRS_model}
  \begin{split}
    y_{i,t} & = (\bh^H_{r,i} \bm \Theta^H \bG +\bh_{d,i}^H ) \bx_t +n_{i,t}\\
    &= (\bH_{r,i} \bm \theta  +\bh_{d,i} )^H \bx_t +n_{i,t},
  \end{split}
\end{equation}
for $ t=1,\ldots, T, ~ i=1,\ldots, K$, where $y_{i,t}$  is the received signal at user $i$ at symbol time $t$; $T$ is the  channel coherence time;
$\bm \Theta =\mbox{Diag}(\bm \theta)$ is the reflecting matrix of the IRS;
$\bx_t\in \Cbb^{N}$ is the transmit signal at the BS;
$\bH_{r,i} =\bG^H\mbox{Diag}(\bh_{r,i})$;
$n_{i,t}\sim {\cal CN} (0,\sigma^2)$ is  circular complex Gaussian noise.
Channel state information (CSI), including the direct channel $\bh_{d,i}$ and the reflected channel $\bG$, $\bh_{r,i}$, is  known at the BS, which may be done via training and estimation of the  cascaded BS-IRS-user CSI; see~\cite{wu2019towards,wang2019channel,yang2020intelligent} for more discussion on the CSI acquisition.

\ifconfver
\begin{figure}[htb!]
  \centering
  \includegraphics[width=0.7\linewidth]{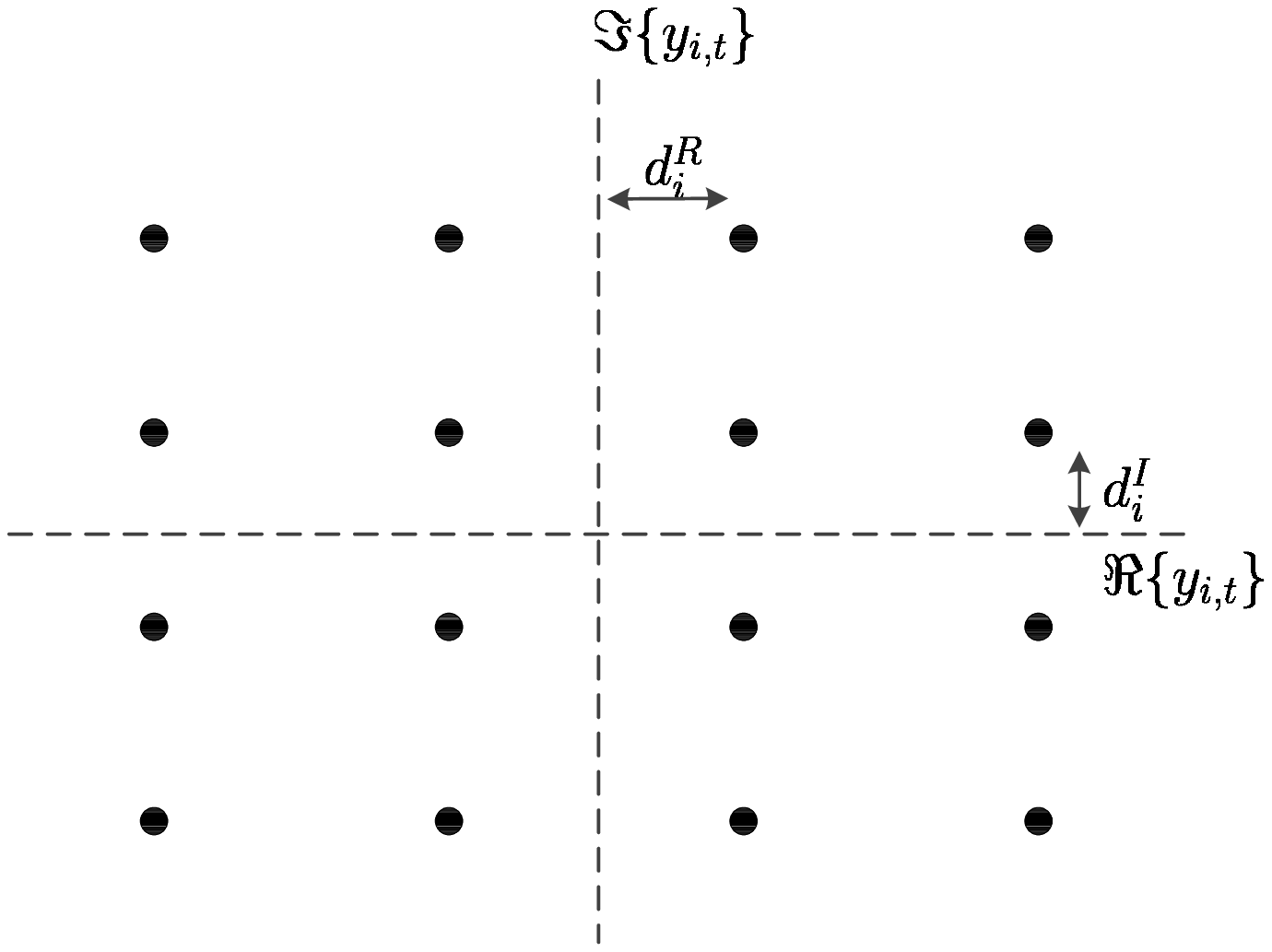}
  \caption{Inter-point spacing $d_{i}^R$ and $d_{i}^I$; $16$-ary QAM.}\label{Fig:d}
\end{figure}
\else
\begin{figure}[htb!]
  \centering
  \includegraphics[width=0.7\linewidth]{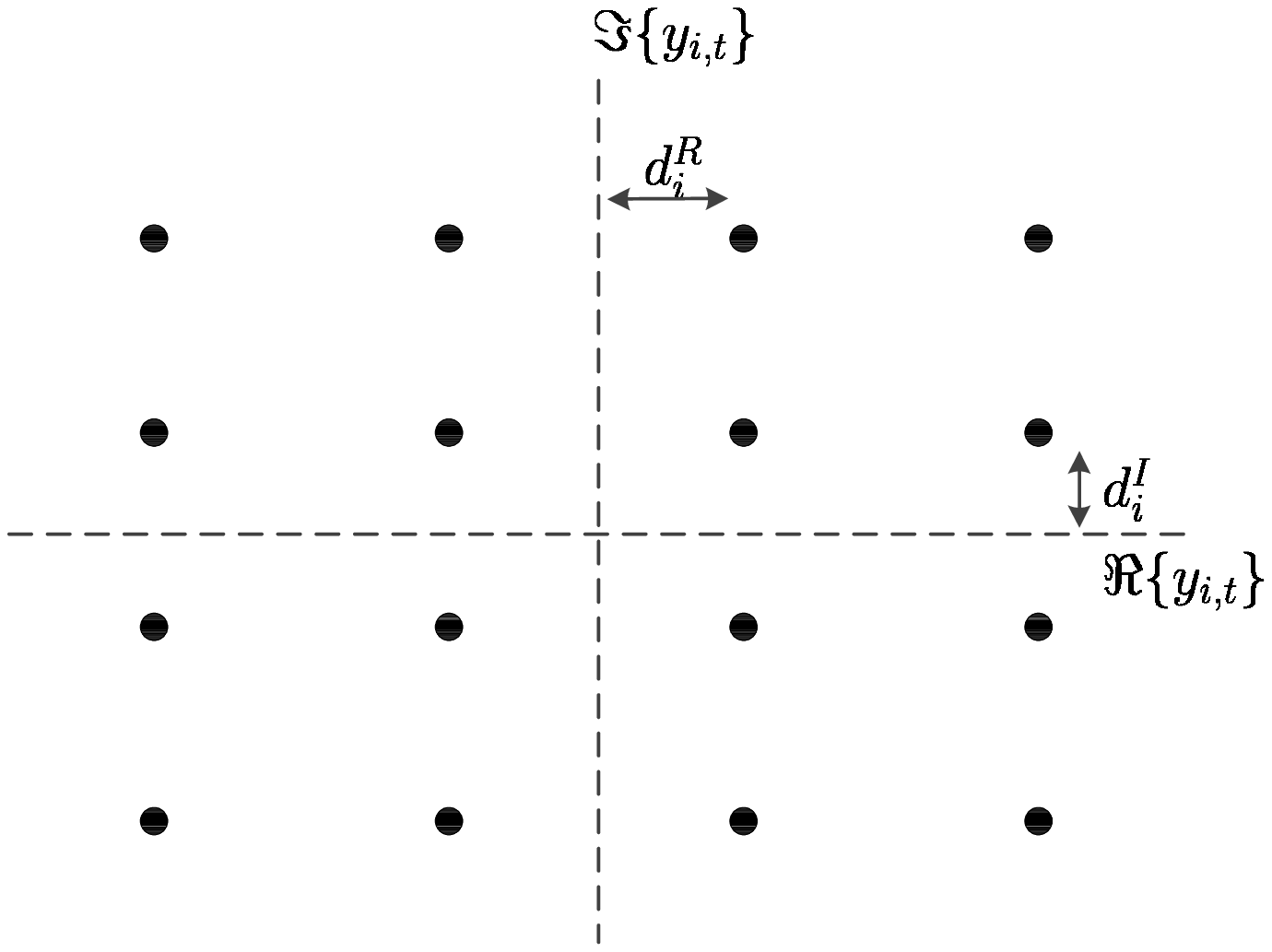}
  \caption{An illustration of the inter-point spacing $d_{i}^R$ and $d_{i}^I$; $16$-ary QAM.}\label{Fig:d}
\end{figure}
\fi

Under the IRS model \eqref{eq:IRS_model}, we aim to jointly design the transmit signal $\bx_t$ and the reflecting coefficient $\bm \theta$ to reduce the SEP of all the users.
Let $s_{i,t}$ be the symbol intended for user $i$ at symbol time $t$. We assume that the symbols are drawn from a QAM constellation, namely,
\[
    s_{i,t} \in \setS_{\sf QAM} \triangleq \{ s_R+ \jj s_I~ | ~ s_R, s_I \in \{ \pm 1, \pm3,\ldots, \pm (2B-1) \} \}
\]
for some positive integer $B$.
The precoder $\bx_t$ and reflecting coefficient $\bm \theta$ are designed such that the users are expected to receive
\[
    y_{i,t} = d_i^R \Re(s_{i,t}) + \jj \cdot d_i^I \Im(s_{i,t}) + n_{i,t},
\]
where $d_i^R>0$ and $d_{i}^I>0$ are the half inter-point spacing  of the real and imaginary parts of the QAM constellation of user $i$, as shown in Fig.~\ref{Fig:d}. 
These inter-point spacings  $d_i^R>0$ and $d_{i}^I>0$  are determined by the BS, and the users are informed of their values during the training phase.
The users detect the symbols by
\[
    \hat{s}_{i,t} = \dec (\Re(y_{i,t})/d_{i}^R) +\jj\cdot \dec(\Im(y_{i,t})/d_i^I),
\]
where $\dec$ is the decision function of $\{ \pm 1, \pm3,\ldots, \pm (2B-1) \}$.
Based on the SEP results in our recent work \cite{shao2019framework}, in the IRS-aided system,  the conditional SEP $\Pr(\hat{s}_{i,t}\neq s_{i,t} ~|~s_{i,t} )$ can be bounded by
\[
    \Pr(\hat{s}_{i,t}\neq s_{i,t} ~|~s_{i,t} ) \leq 2 \max\{ {\sf SEP}_{i,t}^R, {\sf SEP}_{i,t}^I \},
\]
where ${\sf SEP}_{i,t}^R=\Pr(\Re(\hat{s}_{i,t}) \neq \Re(s_{i,t}) ~|~s_{i,t} )$ and 
\ifconfver
\begin{equation}\label{eq:sep_bound}
{\sf SEP}_{i,t}^R \! =\! \begin{cases} \!
Q\! \left(\! \frac{\sqrt{2}b_{i,t}^R}{\sigma} \! \right) \! \!+\! Q \! \left(\! \frac{\sqrt{2}c_{i,t}^R}{\sigma} \! \right), \!\! & \! \! |\Re(s_{i,t})|\! <\! 2B \! -\! 1,\\
Q\left( \frac{\sqrt{2}c_{i,t}^R}{\sigma} \right), & \Re(s_{i,t}) \!= \!2B-1,\\
Q\left( \frac{\sqrt{2}b_{i,t}^R}{\sigma} \right), & \Re(s_{i,t}) \!= \!-2B\!+\! 1
\end{cases} 
\end{equation}
\else
\begin{equation}\label{eq:sep_bound}
{\sf SEP}_{i,t}^R =\! \begin{cases}
Q\left( \frac{\sqrt{2}b_{i,t}^R}{\sigma} \right) \! +\! Q\left(\frac{\sqrt{2}c_{i,t}^R}{\sigma} \right), & |\Re(s_{i,t})| <2B-1,\\
Q\left( \frac{\sqrt{2}c_{i,t}^R}{\sigma} \right), & \Re(s_{i,t}) = 2B-1,\\
Q\left( \frac{\sqrt{2}b_{i,t}^R}{\sigma} \right), & \Re(s_{i,t}) = -2B+1
\end{cases} 
\end{equation}
\fi
with $Q(x)=\int_{x}^{\infty} \frac{1}{\sqrt{2\pi}} e^{-z^2/2} dz$,
\begin{equation*}
  \begin{split}
    b_{i,t}^R = &d_{i}^R - (\Re((\bH_{r,i} \bm \theta  +\bh_{d,i} )^H \bx_t)- d_i^R \Re(s_{i,t})),\\
    c_{i,t}^R = &d_{i}^R + (\Re((\bH_{r,i} \bm \theta  +\bh_{d,i} )^H \bx_t)- d_i^R \Re(s_{i,t})).
  \end{split}
\end{equation*}
The ${\sf SEP}_{i,t}^I$ is defined and characterized in the same way as ${\sf SEP}_{i,t}^R$ by only replacing the ``$R$'' and ``$\Re$'' by ``$I$'' and ``$\Im$'', respectively, in the above results;  Also, $b_{i,t}^I$ and $c_{i,t}^I$ are defined in the same way as $b_{i,t}^R$ and $c_{i,t}^R$, respectively, by  replacing the ``$R$'' and ``$\Re$'' by ``$I$'' and ``$\Im$''; see \cite{shao2019framework} for more details.

Based on the above SEP characterization, our design aims to minimize the SEP in a uniformly fair fashion
\begin{equation}\label{eq:SEP_for}
  \begin{split}
    \min_{\bX, \bd, \bm \theta} &\max_{\substack{  i=1,\ldots, K\\ t=1,\ldots, T}} \max\{  {\sf SEP}_{i,t}^R,{\sf SEP}_{i,t}^I \}\\
    \mbox{s.t.} &~~ \| \bx_t \|^2\leq P, ~~t=1,\ldots, T,~~ \bd\geq \bm 0,\\
    & ~~ |\theta_j|=1, ~~ j=1,\ldots, M,
  \end{split}
\end{equation}
 where $\bX=[\bx_1,\ldots, \bx_T]$, $\bd = [d_{1}^R, \ldots, d_K^R, d_1^I,\ldots d_K^I]^T$, $P$ is the instantaneous total power constraint at the BS.
Here, we also optimize the QAM inter-point spacing for optimized  performance.

 Define the index sets
 \begin{equation*}
\begin{split}
     \Omega^R_{b}  = & \{  (i,t) | \Re(s_{i,t})\neq 2B-1, ~i=1, \ldots, K, ~t=1,\ldots, T \},\\
\Omega^R_{c}  = & \{  (i,t) | \Re(s_{i,t} )\neq \! -2B+1,~i=1, \ldots, K,t=1,\ldots, T \},
\end{split}
 \end{equation*}
 and their indicator function 
 \[
{\cal I}_{i,t}^{R,b} =\begin{cases}
1, & ~ (i,t) \in \Omega_{b}^R\\
0, & ~{\rm otherwise},
\end{cases}
~~{\cal I}_{i,t}^{R,c} =\begin{cases}
1, & ~ (i,t) \in \Omega_{c}^R\\
0, & ~{\rm otherwise}. \end{cases}
\]
 Also, $\Omega^I_{b}$ and $\Omega^I_{c}$ and their index sets ${\cal I}_{i,t}^{I,b}$ and ${\cal I}_{i,t}^{I,c} $ are defined in the same way as  $\Omega^R_{b}$ and $\Omega^R_{c}$, ${\cal I}_{i,t}^{R,b} $ and ${\cal I}_{i,t}^{R,c} $ by only replacing the ``$R$'' and ``$\Re$'' by ``$I$'' and ``$\Im$'', respectively.
 The SEP in \eqref{eq:sep_bound} can be upper bounded by
 \begin{equation}\label{eq:sep_ind}
 \begin{split}
 {\sf SEP}_{i,t}^R \leq \!
2\max \! \left\{ \! {\cal I}_{i,t}^{R,b}\! \cdot \! Q \!\left( \! \frac{\sqrt{2}b_{i,t}^R}{\sigma} \! \right), {\cal I}_{i,t}^{R,c} \! \cdot \! Q \! \left( \! \frac{\sqrt{2}c_{i,t}^R}{\sigma} \! \right) \! \right\}
 \end{split}
\end{equation}
due to $a+b\leq 2 \max\{a,b\}$ for $a,b \geq 0$.
 By substituting the SEP bound \eqref{eq:sep_ind} into problem \eqref{eq:SEP_for}, and by the monotonic decreasing property of  the $Q$ function, problem \eqref{eq:SEP_for} can be explicitly expressed as
 \begin{equation}\label{eq:SEP_final}
   \begin{split}
         \min_{\bX, \bd, \bm \theta} &~ g(\bX, \bd, \bm \theta) \\
    \mbox{s.t.} &~~ \bX\in \setX,\quad \bd\in \setD, \quad \bm \theta \in \Theta,
   \end{split}
 \end{equation}
where
\ifconfver
\begin{equation*}
\begin{split}
& g(\bX, \bd, \bm \theta) \triangleq \\
&\max\{ \!\max_{(i,t)\in \Omega_b^R} \! -b_{i,t}^R,  \! \max_{(i,t)\in \Omega_c^R}  \! -c_{i,t}^R,  \! \max_{(i,t)\in \Omega_b^I}  \! - b_{i,t}^I,  \! \max_{(i,t)\in \Omega_c^I}  \! -c_{i,t}^I \};
\end{split}
\end{equation*} 
\else
\begin{equation*}
\begin{split}
 g(\bX, \bd, \bm \theta) \triangleq 
\max\{ \!\max_{(i,t)\in \Omega_b^R} \! -b_{i,t}^R,  \! \max_{(i,t)\in \Omega_c^R}  \! -c_{i,t}^R,  \! \max_{(i,t)\in \Omega_b^I}  \! - b_{i,t}^I,  \! \max_{(i,t)\in \Omega_c^I}  \! -c_{i,t}^I \};
\end{split}
\end{equation*} 
\fi
  $\setX\triangleq \{\bX~|~\| \bx_t \|^2\leq P, ~t=1,\ldots, T\} $, $ \setD\triangleq \{  \bd\geq \bm 0\}$, $\Theta \triangleq \{ \bm \theta~|~ |\theta_j|=1, ~ j=1,\ldots, M\}$.


\section{Algorithm}
\label{sec:Alg}

Problem \eqref{eq:SEP_final} is non-smooth. To handle this,  we apply smooth approximation to the objective function of \eqref{eq:SEP_final}.
Specifically, we apply the log-sum-exp function to smoothen the $\max$ function, viz.,
\[
  \eta  \log \sum_{i=1}^N e^{x_i/\eta} \approx \max_{i=1,\ldots, N} x_i,
\]
where $\eta>0$ is the smoothing parameter; and the approximation is tight as $\eta\rightarrow 0$. 
This leads to 
\begin{equation}\label{eq:LSE}
  \begin{split}
     \min_{\bX, \bd, \bm \theta} & f(\bX, \bd,\bm \theta) \triangleq \eta \log \Big( \sum_{t=1}^{T} \sum_{i=1}^{K} f_{i,t}(\bX, \bd,\bm \theta) \Big)\\
      \mbox{s.t.} &~~  \bX\in \setX,\quad \bd\in \setD, \quad \bm \theta \in \Theta,
  \end{split}
\end{equation}
where
\ifconfver
\begin{equation*}
\begin{split}
    &f_{i,t} (\bX, \bd,\bm \theta) =\\
    & {\cal I}_{i,t}^{R,b} \cdot e^{-\frac{b_{i,t}^R}{\eta}} + {\cal I}_{i,t}^{R,c}  \cdot e^{-\frac{c_{i,t}^R}{\eta}}+ {\cal I}_{i,t}^{I,b} \cdot  e^{-\frac{b_{i,t}^I}{\eta}}
   + {\cal I}_{i,t}^{I,c} \cdot  e^{-\frac{c_{i,t}^I}{\eta}}.
\end{split}
\end{equation*}
\else
\begin{equation*}
\begin{split}
    f_{i,t} (\bX, \bd,\bm \theta) =
     {\cal I}_{i,t}^{R,b} \cdot e^{-\frac{b_{i,t}^R}{\eta}} + {\cal I}_{i,t}^{R,c}  \cdot e^{-\frac{c_{i,t}^R}{\eta}}+ {\cal I}_{i,t}^{I,b} \cdot  e^{-\frac{b_{i,t}^I}{\eta}}
   + {\cal I}_{i,t}^{I,c} \cdot  e^{-\frac{c_{i,t}^I}{\eta}}.
\end{split}
\end{equation*}
\fi

Problem \eqref{eq:LSE} is non-convex and the non-convexity lies in both the coupling of $\bm \theta$ and $\bx_t$ in the objective function and the unit-modulus constraint of $\theta_i$.
  We apply alternating minimization to handle problem \eqref{eq:LSE} with respect to (w.r.t.) $(\bX,\bd)$  and $\bm \theta$. We apply Nesterov's accelerated projected gradient (APG) method to the $(\bX,\bd)$  and $\bm \theta$ subproblems.
The algorithm sketch is shown in Algorithm~\ref{Al:AM}.
There, ${\sf APG}_{(\bX,\bd)}(\bm \theta)$ stands for the APG solver for solving problem~\eqref{eq:LSE} w.r.t. $(\bX,\bd)$ given $\bm \theta$,
and ${\sf APG}_{\bm \theta} (\bX, \bd)$ the APG solver for solving problem~\eqref{eq:LSE} w.r.t. $\bm \theta$ given $(\bX, \bd)$.
It can be shown that the alternating minimization in Step 5 and 6 yields a non-increasing objective value $f$. Moreover, since the objective value $f$ in \eqref{eq:LSE} is lower bounded, Algorithm~\ref{Al:AM} is guaranteed to converge according to the monotone convergence theorem.

\begin{algorithm}[H]
	\caption{Alternating Minimization for Problem \eqref{eq:LSE}}
	\begin{algorithmic}[1]
		\STATE
		{\bf given}  starting point $(\bX^0,\bd^0,\bm \theta^0)$, smoothing parameter $\eta$.

		\STATE $k= 0$.
	\REPEAT
\STATE  $k=k+1$;
		\STATE Update $(\bX^{k},\bd^{k}) = {\sf APG}_{(\bX,\bd)}(\bm \theta^{k-1})$;

		\STATE Update $\bm \theta^{k} = {\sf APG}_{\bm \theta}(\bX^k, \bd^k)$;

		\UNTIL some stopping criterion is met.
	\end{algorithmic}\label{Al:AM}
\end{algorithm}
~\\[-1cm]

Next, we specify the APG algorithms used in Steps 5 and 6 in Algorithm \ref{Al:AM}.

\subsection{$(\bX,\bd)$ Update}

For any given $\bm\theta$, the subproblem w.r.t. $(\bX, \bd)$ is convex and smooth. We apply APG to solve it. Specifically, the APG iteration is given by
\begin{equation*}
  \begin{split}
    \bX^{j+1} = & \Pi_{\setX} \left( \bZ_X^j -\frac{1}{\beta_j} \nabla_{\bX} f(\bZ_X^j,\bz_d^j,\bm \theta) \right),\\
        \bd^{j+1} = & \Pi_{\setD} \left( \bz_d^j -\frac{1}{\beta_j} \nabla_{\bd} f(\bZ_X^j,\bz_d^j,\bm \theta) \right),
  \end{split}
\end{equation*}
where
\begin{equation*}
  \begin{split}
    \bZ_X^j = & \bX^{j} + \alpha_j (\bX^{j}-\bX^{j-1}),\\
     \bz_d^j = & \bd^{j} + \alpha_j (\bd^{j}-\bd^{j-1}),
  \end{split}
\end{equation*}
with
\begin{equation}\label{eq:FISTA_par}
 \alpha_j= \frac{\xi_{j-1}-1}{\xi_j},\quad \xi_j =\frac{1+\sqrt{1+4\xi_{i-1}^2}}{2};
\end{equation}
$\beta_j$ is chosen by the backtracking line search \cite{beck2009fast};
$\Pi_{\setX}(\bx)\triangleq \arg\min\limits_{\by \in \setX} \| \bx-\by \|^2$  denotes the projection of $\bx$ onto the set $\setX$; note that,
\begin{equation*}
  \begin{split}
   \hat{\bX} =\Pi_{\setX} (\bX) & \Leftrightarrow \hat{\bx}_t=\begin{cases}
                         \bx_t, & \mbox{if } \| \bx_t \|^2\leq P \\
                         \sqrt{P} \frac{\bx_t}{\| \bx_t \|}, & \mbox{otherwise}
                       \end{cases},\quad \forall t;\\
   \Pi_{\setD}(\bd) & =\max\{\bm 0, \bd\}.
  \end{split}
\end{equation*}
The expressions of the gradient $\nabla_{\bX} f$ and $\nabla_{\bd} f$ are lengthy, and owing to the page limit we relegate them to the supplementary material of this paper.
The APG iterations can be very efficient when the constraint set is easy to project, which is true in our case.
 Also, for convex problems, APG is theoretically shown to guarantee a faster convergence rate ${\cal O}(1/j^2)$ than the classic projected gradient algorithm ${\cal O}(1/j)$~\cite{beck2009fast}.

\subsection{$\bm \theta$ Update}

We apply the APG method to the subproblem of optimizing  $\bm \theta$ given $(\bX,\bd)$. The APG update for $\bm \theta$ is given by
\[
    \bm \theta^{j+1} = \Pi_{\Theta} \left( \bz_{\theta}^j - \frac{1}{\gamma_j} \nabla_{\bm\theta} f( \bX, \bd, \bz_{\theta}^j) \right),
\]
where
\[
    \bz_{\theta}^j = \bm \theta^{j} +\alpha_j (\bm \theta^{j} -\bm \theta^{j-1} );
\]
$\alpha_j$ follows the same rule in \eqref{eq:FISTA_par};
$\gamma_j$ is chosen by the backtracking line search~\cite{beck2009fast};
the projection $\Pi_{\Theta}$ takes the form
\begin{equation*}
  \hat{\bm\theta} =\Pi_{\Theta} (\bm\theta) \Leftrightarrow \hat{\theta}_i =\begin{cases}
     \theta_i/|\theta_i|, & \mbox{if } \theta_i \neq 0 \\
     1, & \mbox{otherwise}.
   \end{cases}
\end{equation*}
The expression of the gradient $\nabla_{\bm\theta} f$ is relegated to  the supplementary material of this paper.
Note that the $\bm \theta$ subproblem is non-convex due to the non-convexity of the constraint $|\theta_i|=1$, $i=1,\ldots, M$.
Though it has not been proven that the APG method can achieve faster convergence rate than the projected gradient for non-convex problems, our numerical experience suggests that the acceleration effect of APG is conspicuous.

\section{The  PSK Case}

The design discipline developed above is also applicable to the PSK constellation.
This section will highlight the key idea of how to design the SLP and reflecting coefficients for the PSK constellation.

Suppose the symbols $s_{i,t}$ are drawn from a $L$-ary PSK constellation
\[
    s_{i,t } \in \setS_{\sf PSK} \triangleq \{s ~|~ s=e^{\jj n \frac{2\pi}{L}}, n=1,\ldots, L-1 \}.
\]
By extending the SEP result in our recent work \cite{shao2018multiuser,shao2019onebit}, the  multiuser SEP minimization problem  for the PSK constellations can be formulated as
\begin{equation}\label{eq:SEP_min_PSK}
  \begin{split}
    \min_{\bX,\bm \theta}& \max_{\substack{  i=1,\ldots, K\\ t=1,\ldots, T}} -  a_{i,t}\\
   \mbox{s.t.}&~~ \bX\in \setX, \quad \bm \theta \in \Theta,
  \end{split}
\end{equation}
where
\begin{equation*}
\begin{split}
   a_{i,t}  = & \Re \left( (\bH_{r,i} \bm \theta  +\bh_{d,i} )^H \bx_t s_{k,t}^* \right)\\
   & -
    |\Im \left(  (\bH_{r,i} \bm \theta  +\bh_{d,i} )^H \bx_t s_{k,t}^* \right)| \cot(\pi/L).
    \end{split}
\end{equation*}
Problem \eqref{eq:SEP_min_PSK} can be expressed as
\begin{equation}\label{eq:SEP_PSK2}
  \begin{split}
        \min_{\bX,\bm \theta}& \max_{\substack{  i=1,\ldots, K\\ t=1,\ldots, T}} \max\{ - u_{i,t}, -v_{i,t} \} \\
   \mbox{s.t.}&~~ \bX\in \setX, \quad \bm \theta \in \Theta,
  \end{split}
\end{equation}
where
\ifconfver
\begin{equation*}
  \begin{split}
    u_{i,t} =& \Re \left( (\bH_{r,i} \bm \theta  +\bh_{d,i} )^H \bx_t s_{k,t}^* \right)\\
 &  - \Im \left(  (\bH_{r,i} \bm \theta  +\bh_{d,i} )^H \bx_t s_{k,t}^* \right) \cot(\pi/L),\\
        v_{i,t} =&  \Re \left( (\bH_{r,i} \bm \theta  +\bh_{d,i} )^H \bx_t s_{k,t}^* \right)\\
  & + \Im \left(  (\bH_{r,i} \bm \theta  +\bh_{d,i} )^H \bx_t s_{k,t}^* \right) \cot(\pi/L).
  \end{split}
\end{equation*}
\else
\begin{equation*}
  \begin{split}
    u_{i,t} =& \Re \left( (\bH_{r,i} \bm \theta  +\bh_{d,i} )^H \bx_t s_{k,t}^* \right)- \Im \left(  (\bH_{r,i} \bm \theta  +\bh_{d,i} )^H \bx_t s_{k,t}^* \right) \cot(\pi/L),\\
        v_{i,t} =&  \Re \left( (\bH_{r,i} \bm \theta  +\bh_{d,i} )^H \bx_t s_{k,t}^* \right) + \Im \left(  (\bH_{r,i} \bm \theta  +\bh_{d,i} )^H \bx_t s_{k,t}^* \right) \cot(\pi/L).
  \end{split}
\end{equation*}
\fi
Problem \eqref{eq:SEP_PSK2} takes a similar and even simpler form than the QAM case \eqref{eq:SEP_final}, and we can use the exactly same smoothing and APG tricks in Sec.~\ref{sec:Alg} to tackle it.
We shall omit the algorithm details  here due to space limitation.

\section{Simulation Results}
In this section, we test the performance of the proposed SLP design. For convenience, we will call the proposed design ``SLP-IRS''.
The simulation settings are as follows.
The BS has $N=8$ antennas. The  location of the BS is $(0,0)$ and the location of IRS is $(50,0)$.
There are $K=8$ users, randomly lying on a circle centered at $(40,20)$ with radius $10$ m.
The path loss of all the channels follows the model
\[
    L(d) =C_0 \left(  d/D_0 \right)^{-\alpha},
\]
where $C_0$ is the path loss at the reference distance $D_0=1$; $d$ denotes the link distance and $\alpha$ is the path loss exponent.
The BS-IRS channel $\bG$ is given by
\[
    \bG = \beta \bG_{\sf LOS} + \sqrt{1-\beta^2} \bG_{\sf NLOS},
\]
where $\beta$ is the Rician factor; $\beta=1$ and $\beta_0$ correspond to the pure line-of-sight channel and Rayleigh channel, respectively;
the $\bG_{\sf LOS}$  and $\bG_{\sf NLOS}$ denote the line-of-sight and Rayleigh fading channel.
In particular, we have  $\bG_{\sf LOS}=(\frac{1}{\sqrt{2}}+\jj \frac{1}{\sqrt{2}})\bm 1_{M\times N}$, and the elements of $\bG_{\sf NLOS}$   follow the ${\cal CN}(0,1)$ distribution in the i.i.d. fashion~\cite{khalighi2001capacity}.
The channel  $\bh_{r,k}$ and $\bh_{d,k}$ are generated by the same way.
We will use $\alpha_{BI}$, $\alpha_{Iu}$ and $\alpha_{Bu}$ to denote the path loss exponents of the BS-IRS, IRS-user and BS-user channels, respectively.
The $\beta_{BI}$, $\beta_{Iu}$ and $\beta_{Bu}$ denote the Rician factors of the
BS-IRS, IRS-user and BS-user channels, respectively.
In the following simulations, we set $C_0=20$ dB, $\alpha_{BI}=2.2$, $\alpha_{Bu} = \alpha_{Iu} = 2.8$, $\beta_{BI}=0.6$, $\beta_{Bu}=\beta_{Iu} =0$.
The instantaneous total power at the BS is $P=20$ dB. The transmission block length $T=10$.
The parameters in Algorithm~\ref{Al:AM} are as follows: Algorithm \ref{Al:AM} will stop when $k>20$, or when the successive difference satisfies $\| \bX^{k} -\bX^{k-1}\|_F + \| \bd^{k}- \bd^{k-1} \| +\| \bm \theta^{k} -\bm \theta^{k-1} \|\leq 10^{-5} $;
the smoothing parameter is $\eta=0.01$; the APG algorithm for each problem stops when the maximum $1,000$ iteration is met, or when the successive difference satisfies $\|\bX^j-\bX^{j-1}  \|_F+ \| \bd^{k}- \bd^{k-1} \| \leq 10^{-5}$  for QAM ($\|\bX^j-\bX^{j-1}  \|_F\leq 10^{-5}$ for PSK) and $\| \bm \theta^{k} -\bm \theta^{k-1} \|\leq 10^{-5}$.

We consider two benchmark schemes:
(1) zero-forcing (ZF) precoding under average power constraint and without IRS;
(2) optimal linear beamforming with IRS (OLB-IRS) \cite{wu2019intelligent};
(3) SLP  without   IRS.
We adopt bit-error rate (BER) as the performance metric. All the results were averaged over $5,000$ independent channel realizations.
\ifconfver
 \begin{figure}[htb!]
  \centering
  \includegraphics[width=0.8\linewidth]{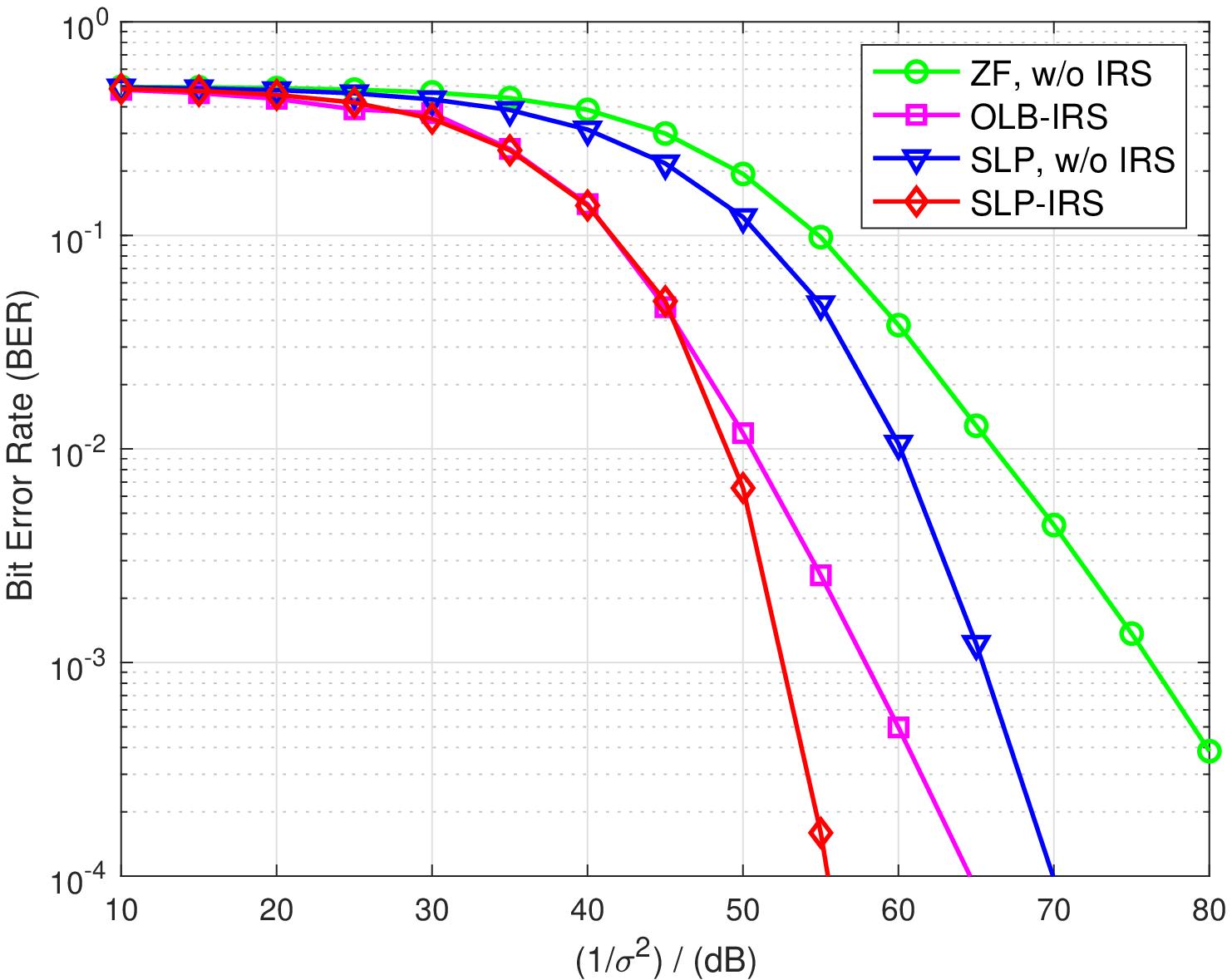}
  \caption{BER performance versus  noise power; $16$-ary QAM.}\label{fig:IRS_16QAM}
\end{figure}
\else
 \begin{figure}[htb!]
  \centering
  \includegraphics[width=0.6\linewidth]{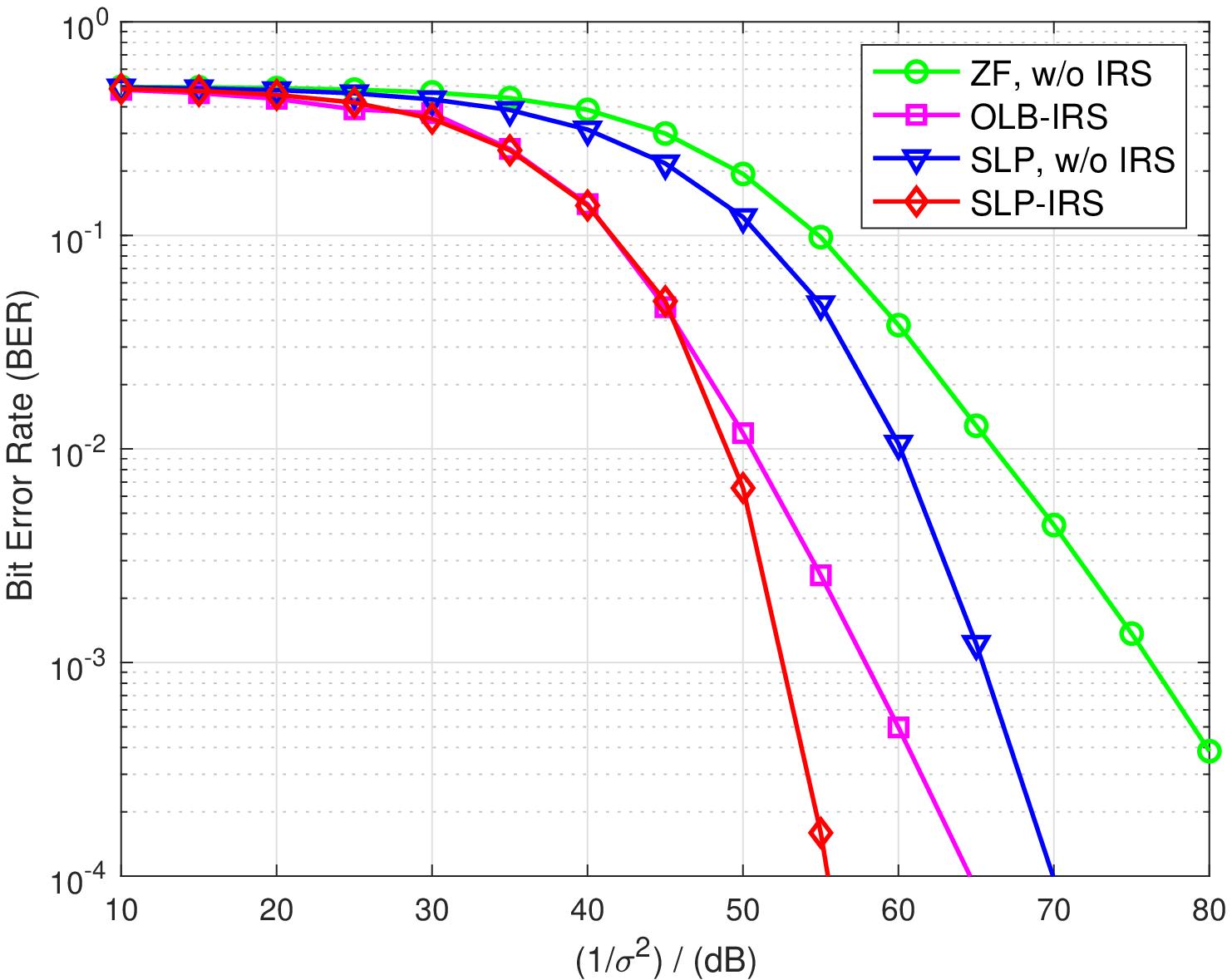}
  \caption{BER performance versus  noise power; $16$-ary QAM.}\label{fig:IRS_16QAM}
\end{figure}
\fi

\ifconfver
 \begin{figure}[htb!]
  \centering
  \includegraphics[width=0.8\linewidth]{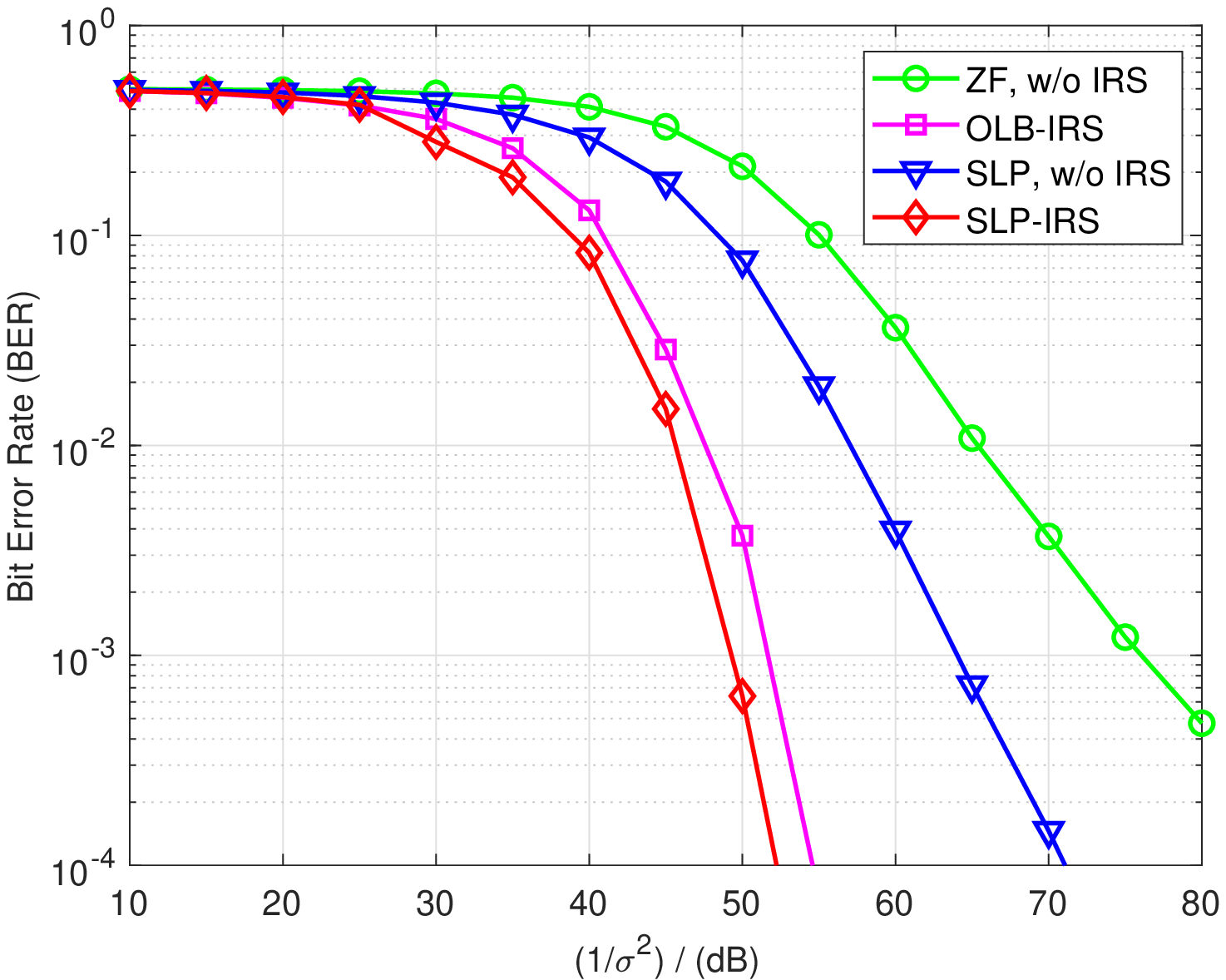}
  \caption{BER performance versus  noise power; $8$-ary PSK.}\label{fig:IRS_PSK}
\end{figure}
\else
 \begin{figure}[htb!]
  \centering
  \includegraphics[width=0.6\linewidth]{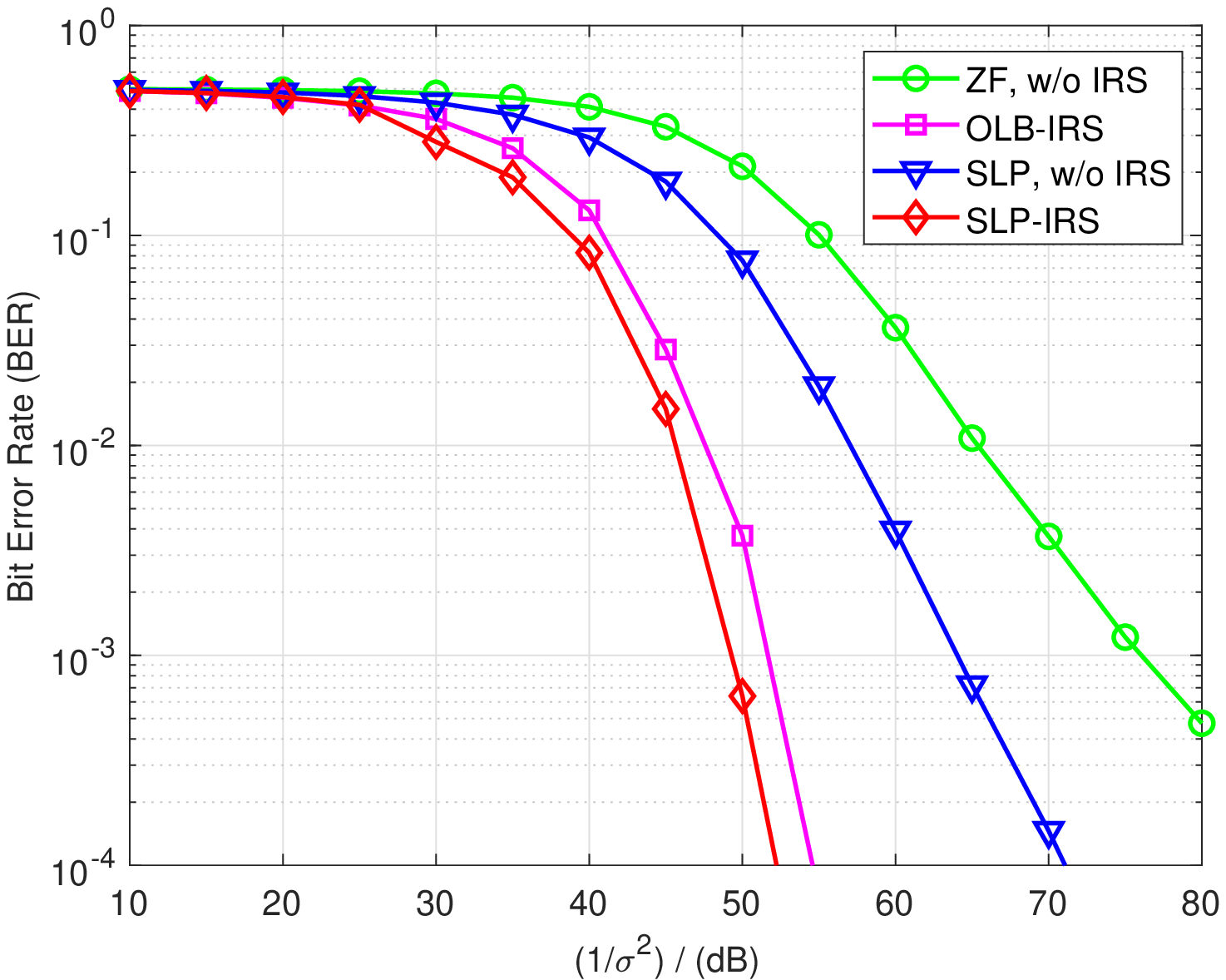}
  \caption{BER performance versus  noise power; $8$-ary PSK.}\label{fig:IRS_PSK}
\end{figure}
\fi

Figs.~\ref{fig:IRS_16QAM} and \ref{fig:IRS_PSK} show the BER performance w.r.t. the noise power $\sigma^2$ for $16$-ary QAM and $8$-ary PSK, respectively.
The number of IRS elements is $M=32$.
It is seen that under the aid of IRS, SLP-IRS outperforms the OLB-IRS.
Also, the IRS-aided SLP performs much better than the SLP without the IRS.
In particular, the gap between the SLP designs with and without the aid of IRS in this case is more than $10$ dB at the BER level $10^{-3}$ for both QAM and PSK signaling.

\ifconfver
 \begin{figure}[htb!]
  \centering
  \includegraphics[width=0.8\linewidth]{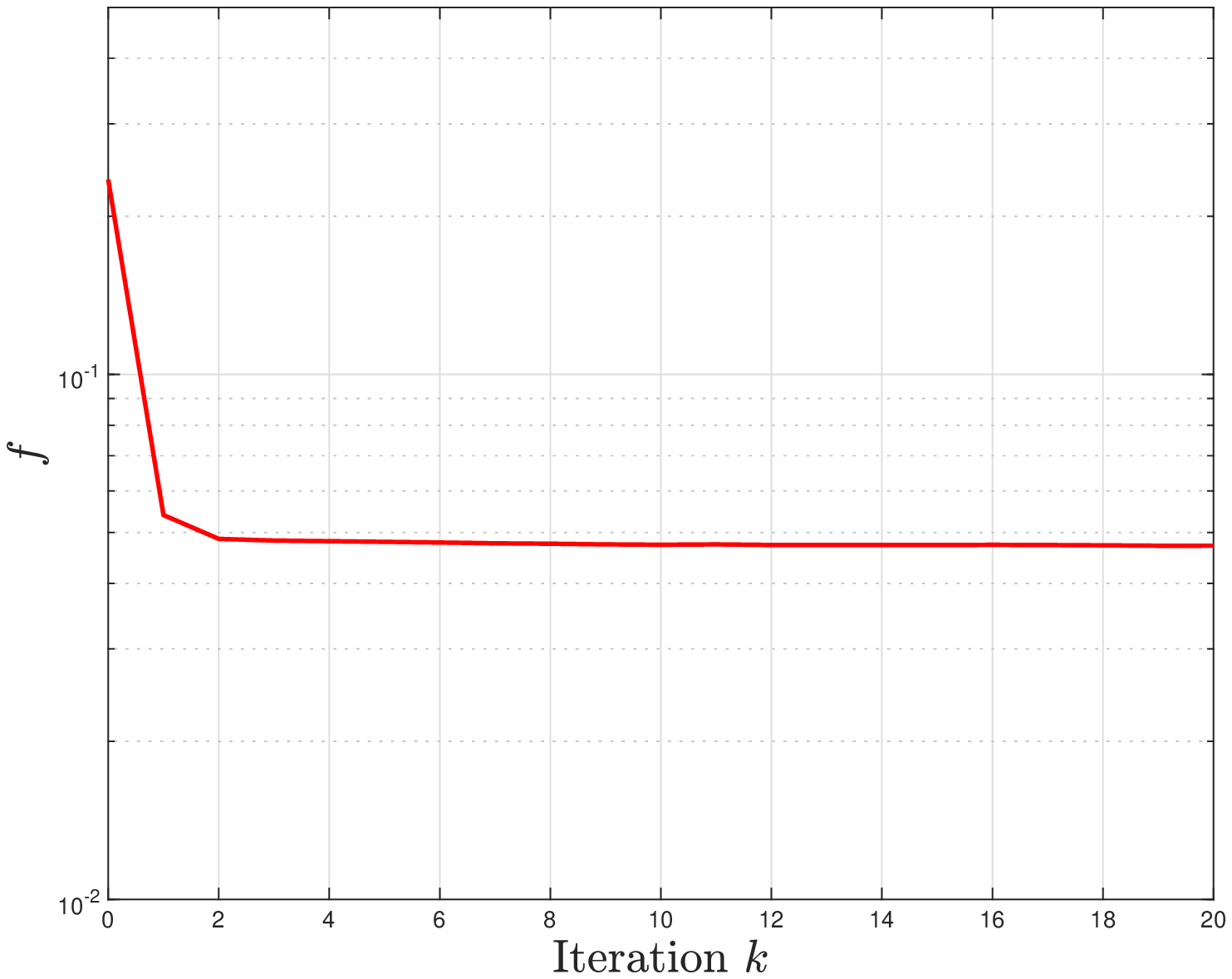}
  \caption{Convergence of function value $f$ with respect to the iteration number $k$; $16$-ary QAM.}\label{fig:func_val}
\end{figure}
\else
\begin{figure}[htb!]
  \centering
  \includegraphics[width=0.6\linewidth]{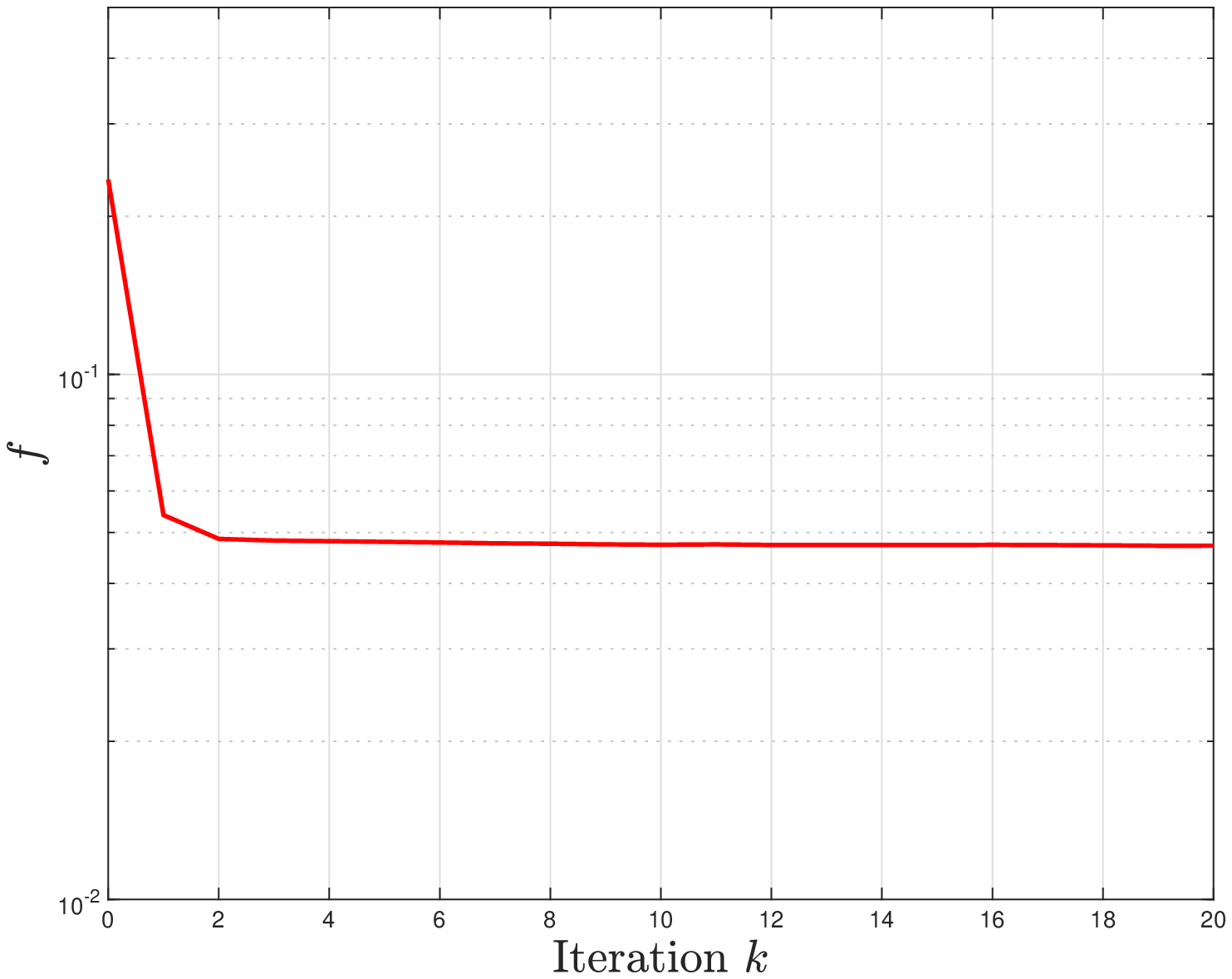}
  \caption{Convergence of function value $f$ with respect to the iteration number $k$; $16$-ary QAM.}\label{fig:func_val}
\end{figure}
\fi

We test the convergence property of Algorithm~\ref{Al:AM} in Fig.~\ref{fig:func_val}. We show the convergence behavior of the function value $f$ in problem~\eqref{eq:LSE} with respect to the iteration number $k$. The simulation settings are the same as those in Fig.~\ref{fig:IRS_16QAM}. 
It is seen that Algorithm~\ref{Al:AM} converges within a few iterations.

\begin{figure}[htb!]
  \centering
  \begin{subfigure}[b]{0.49\linewidth}
  \includegraphics[width=\linewidth]{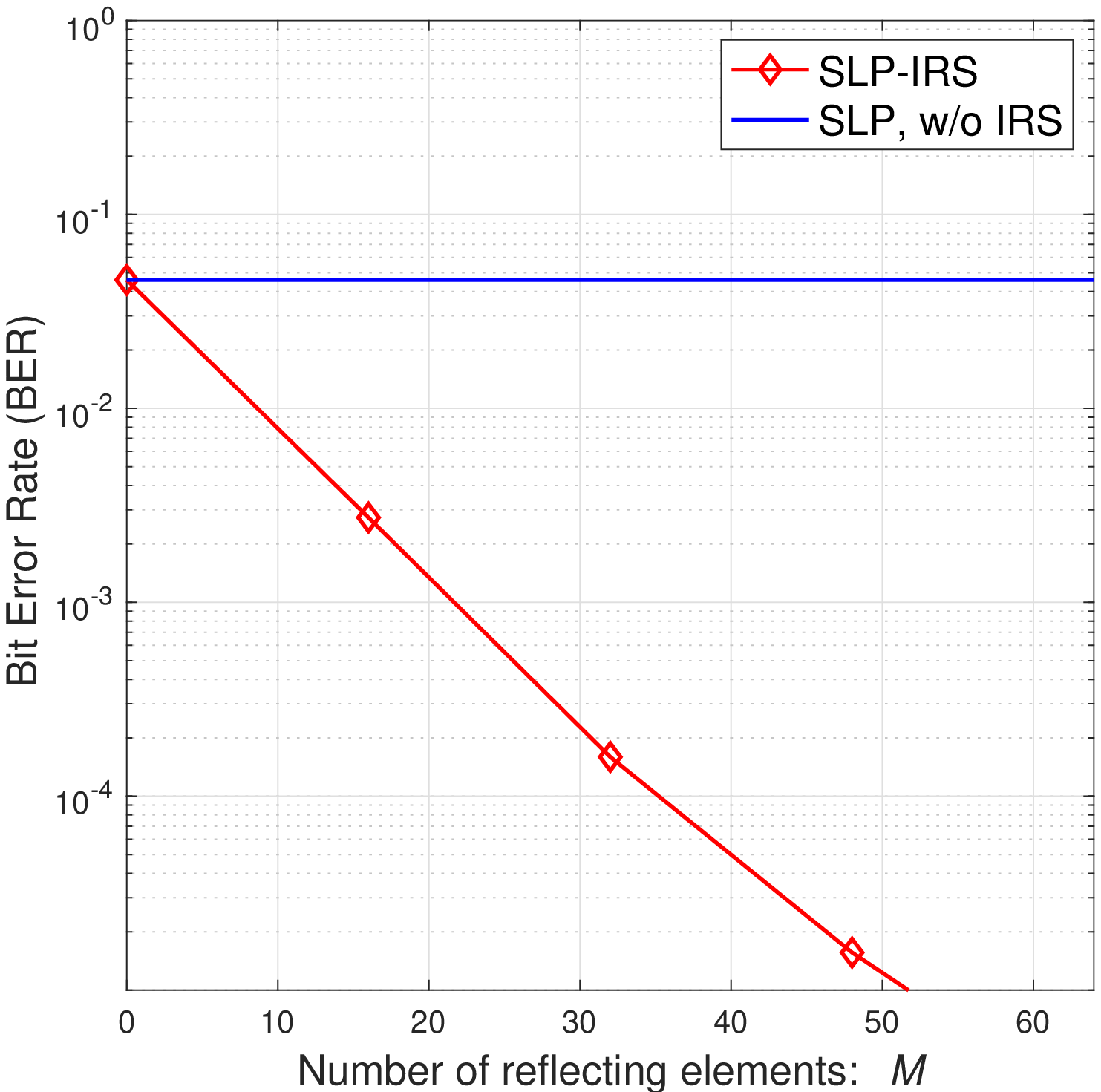}
    \caption{$16$-ary QAM}\label{fig:M_QAM}
  \end{subfigure}
  \begin{subfigure}[b]{0.49\linewidth}
  \includegraphics[width=\linewidth]{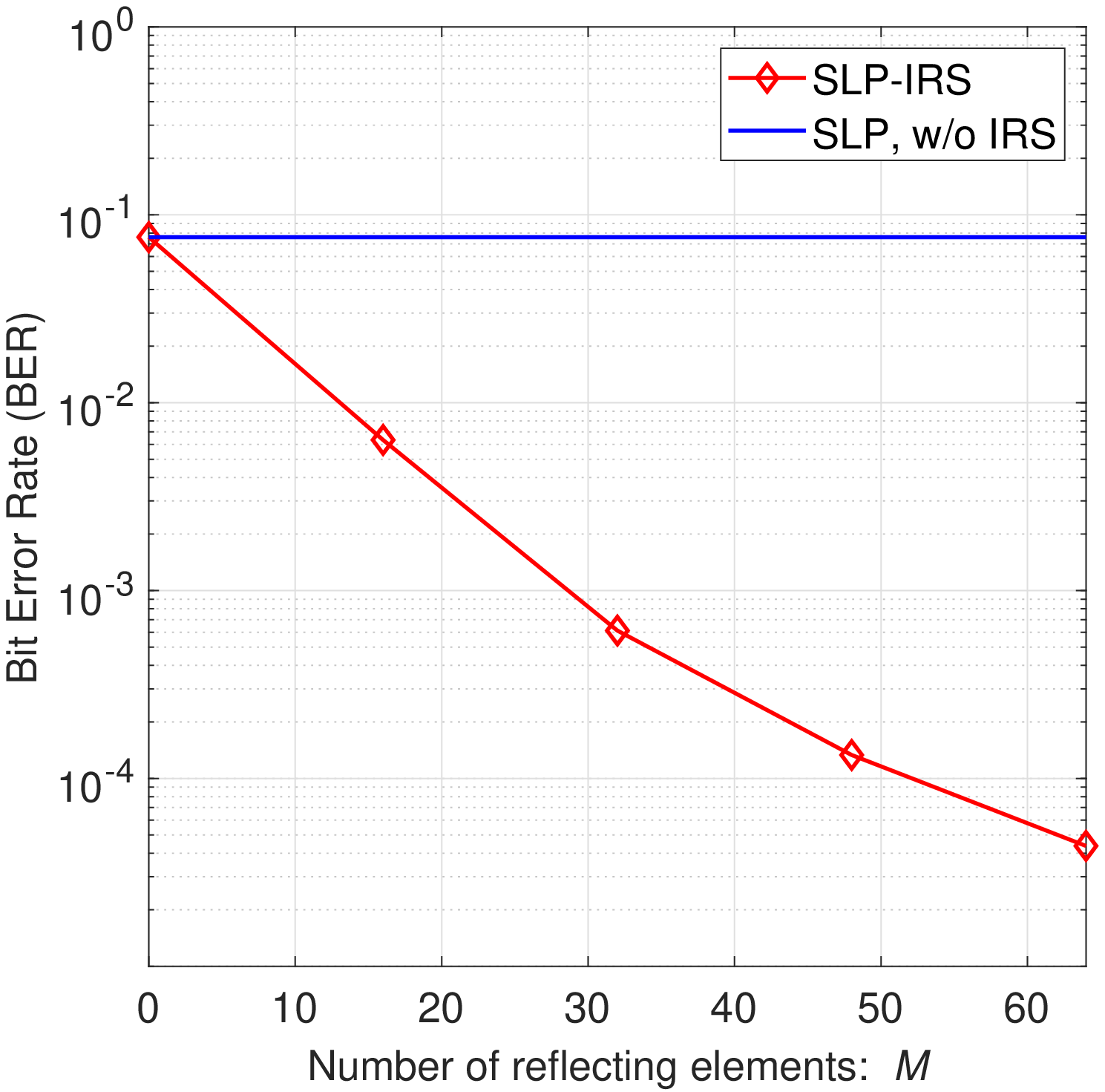}
  \caption{$8$-ary PSK}\label{fig:M_PSK}
  \end{subfigure}
  \caption{BER performance versus number of IRS elements.}\label{fig:M}
\end{figure}

Fig.~\ref{fig:M} shows the BER performance for different number of IRS elements. Fig.~\ref{fig:M}(a) shows the result for $16$-ary QAM at the noise power level  $-55$ dB; Fig.~\ref{fig:M}(b) shows the result $8$-ary PSK at the noise power level $-50$ dB. We  increase the number of IRS elements from $0$ to $64$. It is seen that the BER performance of SLP-IRS increases as the number of IRS elements increases. And the performance improvement is significant when the number of IRS elements is large.

\section{Conclusion}

In this letter, we have investigated a  minimum-SEP SLP design with the aid of IRS
for multiuser MISO downlink.
We consider both the QAM and PSK constellation cases, and we tackle the design formulation by alternately optimizing the precoder and IRS angle.
Simulation results have demonstrated that, with the aid of IRS, the BER performance of the proposed SLP design is much better than the  SLP without  IRS, especially when the size of IRS is large.
We should mention that this work assumes IRS has fixed amplitude and continuous phase. As a future work, it is worthwhile to consider IRS with discrete phase and/or adjustable amplitude. 
%



\input{IRS_WCL_final_arxiv.bbl}

\end{document}